\documentclass[a4paper]{article}

\usepackage{graphicx}
\usepackage{amsmath}
\usepackage{subcaption}
\usepackage{xcolor}
\usepackage{listings}

\usepackage{url}

\definecolor{codegreen}{rgb}{0,0.6,0}
\definecolor{codegray}{rgb}{0.5,0.5,0.5}
\definecolor{codepurple}{rgb}{0.58,0,0.82}
\definecolor{backcolour}{rgb}{0.95,0.95,0.92}
 
\lstdefinestyle{mystyle}{
    backgroundcolor=\color{backcolour},   
    commentstyle=\color{codegreen},
    keywordstyle=\color{magenta},
    numberstyle=\tiny\color{codegray},
    stringstyle=\color{codepurple},
    basicstyle=\footnotesize,
    breakatwhitespace=false,         
    breaklines=true,                 
    captionpos=b,                    
    keepspaces=true,                 
    numbers=left,                    
    numbersep=5pt,                  
    showspaces=false,                
    showstringspaces=false,
    showtabs=false,                  
    tabsize=2
}
 
\newcommand{\preprinttitle}{Structify-Net: Random Graph generation with controlled size and customized structure}

\newcommand{\listauthors}{\raggedright 
Rémy Cazabet\textsuperscript{1}, \space
Salvatore Citraro\textsuperscript{2},
Giulio Rossetti\textsuperscript{2}
}

\newcommand{\listinstitutions}{
\textsuperscript{1} Univ Lyon, UCBL, CNRS, INSA Lyon, LIRIS, UMR5205, F-69622 Villeurbanne, France 
\\
\textsuperscript{2} Institute of Information Science and Technologies “A. Faedo” (ISTI), National Research Council (CNR), Italy
}

\newcommand{\email}{remy.cazabet@univ-lyon1.fr}

\newcommand{\preprintabstract}{Network structure is often considered one of the most important features of a network, and various models exist to generate graphs having one of the most studied types of structures, such as blocks/communities or spatial structures. In this article, we introduce a framework for the generation of random graphs with a controlled size ---number of nodes, edges--- and a customizable structure, beyond blocks and spatial ones, based on node-pair rank and a tunable probability function allowing to control the amount of randomness. We introduce a \textit{structure zoo} ---a collection of original network structures--- and conduct experiments on the small-world properties of networks generated by those structures. Finally, we introduce an implementation as a Python library named \textit{Structify-net}. }

\newcommand{\preprintkeywords}{Network Generation, Random Graphs, Network Structure, Python Library}

\usepackage{tikz}

\usepackage[margin=1in,marginparwidth=4.2cm,marginparsep=0.5cm]{geometry}

\usepackage{marginnote}
\usepackage{pdflscape}
\usepackage{ifthen}
\reversemarginpar  
\usepackage{lipsum} 
\usepackage{calc}
\usepackage{siunitx}
\usepackage{titlesec}
\usepackage{indentfirst}
\usepackage[utf8]{inputenc}
\usepackage[T1]{fontenc}
\usepackage[
backend=biber,
natbib=true,
sortcites=true,
defernumbers=true,
style=authoryear,
citestyle=authoryear-comp,
maxnames=99,
maxcitenames=2,
uniquename=init,
giveninits=true,
terseinits=true,
url=false
]{biblatex}
\DeclareNameAlias{default}{family-given}
\renewbibmacro{in:}{%
  \ifentrytype{article}{}{\printtext{\bibstring{in}\intitlepunct}}} 
\DeclareFieldFormat[article]{pages}{#1} 
\AtEveryBibitem{\ifentrytype{article}{\clearfield{number}}{}} 
\DeclareFieldFormat[article, inbook, incollection, inproceedings, misc, thesis, unpublished]{title}{#1} 
\usepackage{csquotes}
\RequirePackage[english]{babel} 
\DeclareBibliographyCategory{ignore}
\addtocategory{ignore}{recommendation} 
\usepackage{nameref}
\DeclareFieldFormat{doi}{\url{https://doi.org/#1}}

\usepackage{graphbox}  
\usepackage{microtype}
\DisableLigatures[f]{encoding = *, family = * }
\RequirePackage[default,scale=0.90]{opensans}
\usepackage{xcolor}
\definecolor{darkgray}{HTML}{808080}
\definecolor{mediumgray}{HTML}{6D6E70}
\definecolor{ligthgray}{HTML}{d9d9d9}
\definecolor{pciblue}{HTML}{74adca}
\definecolor{opengreen}{HTML}{77933c}
\usepackage{changepage}
\usepackage[aboveskip=1pt,labelfont=bf,labelsep=period,singlelinecheck=off]{caption}

\usepackage[pdfborder={0 0 0},   
    colorlinks=true,
    linkcolor=blue,
    urlcolor=blue,
    citecolor=blue]{hyperref}  

\newcommand{\beginingpreprint}{
\vspace*{0.5cm}
\begin{flushleft}
\baselineskip=0pt

{\Huge
\fontseries{sb}\selectfont{\preprinttitle}}
\end{flushleft}
\vspace*{0.25cm}
\begin{flushleft}
\Large
\listauthors
\end{flushleft}
\bigskip
{\raggedright
\listinstitutions}
\\
\\
\textbf{Correspondence: } \href{mailto:\email}{\email}\\
\\
\vspace*{0.5cm}
\fcolorbox{ligthgray}{ligthgray}{
\parbox{\textwidth - 2\fboxsep}{
\vspace{0.25cm}
\textbf{\large{\textsc{Abstract}}}\\

\preprintabstract\\

\textbf{\emph{Keywords: }}\preprintkeywords

\vspace{0.25cm}}
}
\newpage
\newgeometry{margin=1in}
}

\begin{document}

\begin{tikzpicture}[remember picture,overlay]
\hypersetup{hidelinks}
\node[anchor=north west,yshift=80pt,xshift=-20pt]
{ \href{https://doi.org/10.24072/pci.networksci.100114}{\includegraphics[height=35mm]{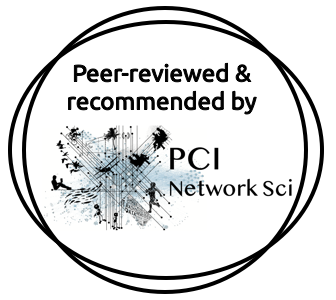}} } ;
\end{tikzpicture}

\beginingpreprint



\section{\centering Introduction}

The structure of networks has long been one of the most studied research questions in network science. In this article, we introduce a method to generate networks of a chosen size, organized according to a structure that can be expressed as an arbitrary ranking function for node pairs. This process thus allows the generation of classic structures such as communities, blocks, and spatial organizations, but also more exotic ones. We subsequently show an application of this framework to study network properties, by extending the classic small-world experiment by (\cite{watts1998collective}). A python library (\cite{structify}) allowing reproduction of the results and generating networks with custom structures is also introduced.

\subsection*{Context}
Generating networks of a chosen size and respecting some constraints is a key topic in network science. It is used in various tasks, for instance, to study network properties (\cite{wang2003complex}), as null models (e.g., \cite{durak2013scalable}), to study the impact on diffusion processes (e.g., \cite{odor2021switchover}), as benchmarks (e.g., \cite{lancichinetti2008benchmark}), etc.  In this article, we focus more particularly on a class of random graph models, in which the probability of observing edges between each pair of nodes is independent of the probability of observing edges between other pairs. This class of models is commonly used to generate various structures, from the homogeneous Erdős–Rényi (ER) generator, to configuration models preserving node degrees, block structures (\cite{abbe2017community}), spatial structures (\cite{waxman1988routing,cazabet2017enhancing}), etc. See section \ref{relatedWork} for an overview of related works.

The originality of our work is to propose a generic framework to generate many different network structures while allowing to set:
\begin{itemize}
\item The number of nodes $n$;
\item The number of edges $m$ (equivalently, the density);
\item A parameter $\epsilon \in [0,1]$ controlling the strength of the structure bias, i.e., the network is fully determined by the structure definition when $\epsilon=0$, and increase in randomness with $\epsilon$, becoming an ER network for $\epsilon=1$. 
\end{itemize}

The main advantage of our proposition compared with existing frameworks such as SBM or latent space models is 1) the simplicity for the user to design their own structure logic, by providing their own ranking function (see Section \ref{method}), 2) to control the \textit{strenght} of the structure bias using a single numerical parameter. Structify-Net is not intended to be used in parameter inference tasks, but only for the generation of null models, reference models, and random graphs with controlled properties in general.

\subsection{Motivational examples}

Generating multiple random networks with common properties, either fixed beforehand or preserved from an observed network, is needed in most domains of network science. In this section, we list three motivational examples of usages of such models, for which Structify-net provides a simple practical solution. These examples are in no way exhaustive since random networks are used in many other contexts in network science.

\subsubsection*{Reference model}
When one is interested in studying a network property, such as the transitivity or the homophily of a node attribute, one usually needs a random graph model as a reference. The simplest one of them is the Erdős–Rényi (ER) random graph model, in which only the nodes and the expected number of edges are preserved; but in most cases, one would like to compare with other reference models. For instance, when studying the transitivity, one might wonder if the observed transitivity of a network is significantly higher than the transitivity of a similar graph having a spatial structure, a block structure, or a strong degree of heterogeneity, etc. To explore those hypotheses, one needs a generative model to generate random networks having the same number of nodes and edges as the observed network, but with a particular structure.

\subsubsection*{Benchmark for network tasks}
A generative model can also be used as a benchmark to test algorithms for complex network tasks developed for capturing fundamental patterns of networks and their functions. A common task where synthetic networks are used to evaluate the performances of an algorithm is community detection, namely the task of identifying ---in its general intuition--- well-connected and/or well-separated groups of nodes within a network (\cite{fortunato2016community}). Generators with planted communities are used to estimate the agreement between their ground-truth structure and the communities captured by an algorithm, as for instance the LFR benchmark (\cite{lancichinetti2008benchmark}). To test the robustness of a wide variety of algorithms defined for different purposes, one needs generators able to model a wide range of planted structures capturing the properties one intends an algorithm to handle. These properties can range from density to homophily, plus any preferred combination of structural properties leading to clique-, grid-, and star-based structures, among others (\cite{yamaguchi2020controlling}).

\subsubsection*{Influence of structure on dynamical processes}
Dynamical processes on networks, such as diffusion or synchronization, have been studied for a long time. Typical examples would be the diffusion of pandemics or polarization on social media. The structure and properties of the network are well-known to have various effects on these processes. For instance, diffusion speed depends on degree heterogeneity (\cite{barthelemy2004velocity}), community structure (\cite{peng2020network}) and clustering (\cite{zhuang2017clustering}), etc. In order to experiment with which factors might control the speed and scale of a particular diffusion process, one should compare such processes with random networks that 1)are comparable in terms of size, i.e., number of nodes and edges, and 2)differ in their structure.

\begin{figure}[h]
\centering
\includegraphics[width=0.98\textwidth]{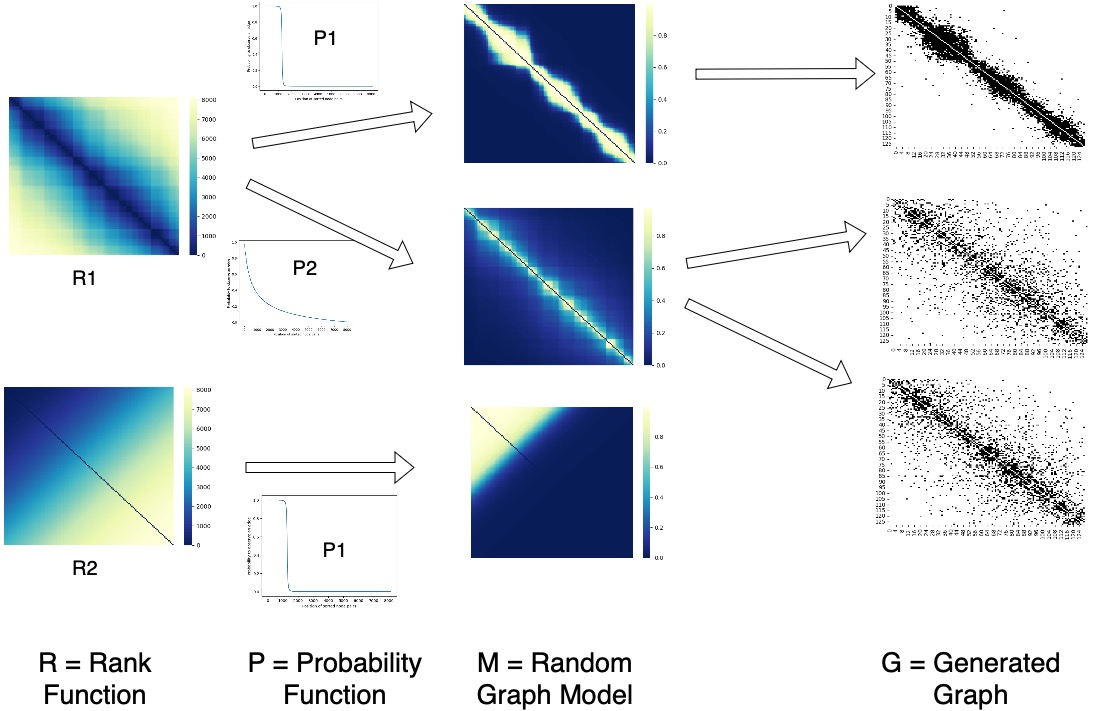}
\caption{Generating networks using the Structify-Net approach. A rank function defines an ordering between node pairs. A probability function is used to assign an edge probability to each node pair based on their rank in the ordering. This gives a Random graph model, that can be used to generate graph instances. Note how the same Rank function $R1$ can give 2 Random Graph Models using different Probability functions $P1$ and $P2$, how the same Probability function $P1$ is used for two different Rank functions $R1$ and $R2$, and how multiple graphs can be generated from a single Random Graph Model.}

\end{figure}

\section{\centering Method}
\label{method}
Structify-Net principle is to create probabilistic random graph generators in two steps:
\begin{enumerate}
    \item A rank function $R$ sets an order among the node pairs, from \textbf{most likely to appear} to \textbf{less likely to appear};
    \item A probability function $P$ assigns to each pair of nodes a probability to be connected by an edge, based on its rank.
\end{enumerate}
$P$ allows to control the expected number of edges $\widehat{m}$. The function $P$ is independent from the graph structure represented by $R$; and reciprocally $R$ is independent from the expected number of edges $\widehat{m}$ or the function $P$.

\subsection{Rank function}
The principle of Structify-Net generator is to describe a network structure by an edge-pair ranking function. More formally, we define $R(u,v)=r$ the function assigning a value $r$ to each undirected node pair, such as $r\in[1,\frac{n*(n-1)}{2}]$ corresponds to the rank of the node pair $(u,v)$, and $r(u,v)<r(u2,v2)$ means that it is more likely to observe an edge between the pair $(u,v)$ than between the pair $(u2,v2)$. This function can be expressed directly in that form, or be trivially derived from a function $R'$ assigning a cost to each node pair, coupled with a sorting function, ranking pairs by increasing or decreasing values of $R'(u,v)$. In practice, in that case, we also add an infinitesimal random value $\iota$ to each cost, in order to avoid ties.

An intuitive example of such a cost function is for the \textit{spatial} structure: given a position vector $W_u$ for each node $u$ (provided to mimic real data, or generated in fully synthetic network generation), the tendency to observe edges can be a function of the distance, e.g., $R'(u,v)=||W_u,W_v||$. By sorting node pairs by increasing distance, we obtain a spatial structure such that the closer the nodes, the higher their tendency to be connected.

Section \ref{zoo} describes in more detail various types of network structures that can be represented this way.

\subsection{Probability Function}
To go from a ranking of node pairs to a random network generator, we use a rank probability function assigning a probability to each rank, i.e., $P(r)\in[0,1]$. The only constraint to this function is that it must be non-increasing, so that a node pair of lower rank is at least as likely to be connected by an edge than a node pair of higher rank. 

The probability function controls the expected number of edges: 
\[
\widehat{m}=\sum_r^L P(r)
\]
with $L=\frac{n(n-1)}{2}$

\begin{figure}[h]
\centering
\begin{subfigure}{.45\textwidth}
  \centering
  \includegraphics[width=.85\linewidth]{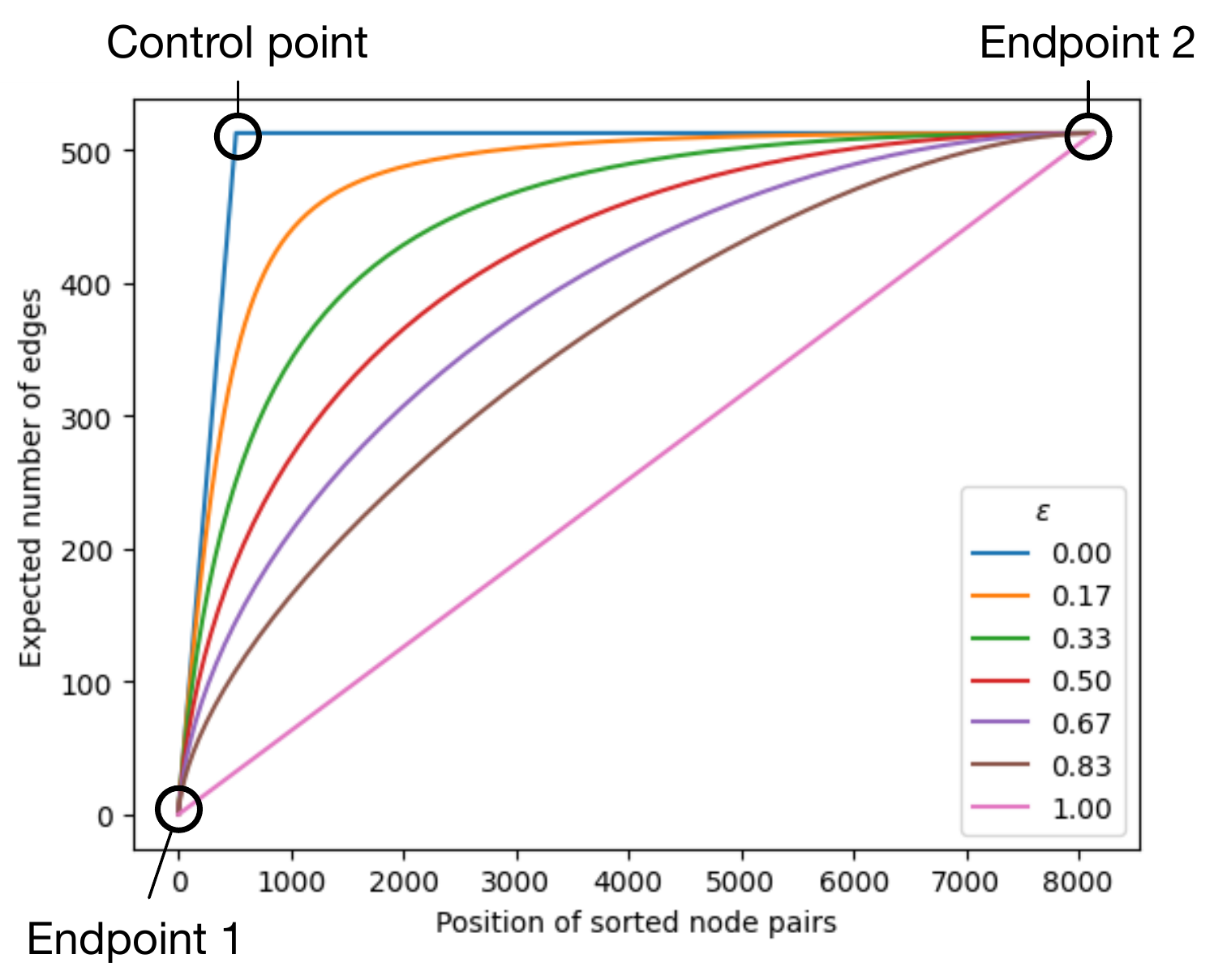}
  \caption{Bézier interpolation of the number of edges encountered at a given rank}
  \label{fig:sub1}
\end{subfigure}%
\hspace{1em}
\begin{subfigure}{.45\textwidth}
  \centering
  \includegraphics[width=.93\linewidth]{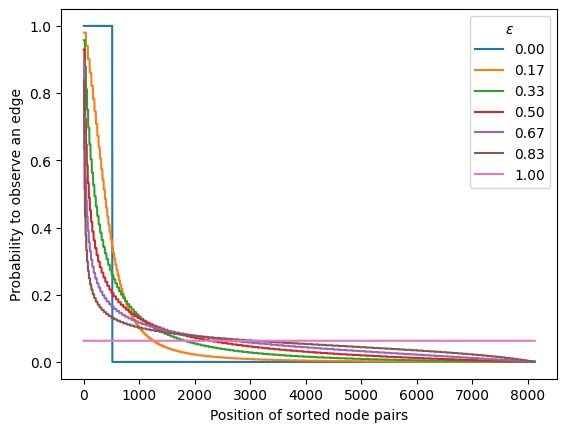}
  \caption{The corresponding probability function, i.e., probability of observing an edge at a given rank}
  \label{fig:sub2}
\end{subfigure}
\caption{Example of probability functions for various values of $\epsilon$. In this example, we set $m=128,n=512$}
\label{fig:bezierP}
\end{figure}

\subsubsection*{Bézier Interpolated Probability Function}
We propose a family of probability functions controlled by 1) the target expected number of edges $m$, 2) a parameter $\epsilon$, which controls how strongly is the random graph driven by the planted structure. The family is defined as follows: at one extreme ($\epsilon=1$), the probability of observing an edge is independent of the rank, i.e., $P(r)=m/L$, as in an ER random graph. Conversely, at the other extreme $\epsilon=0$, the $m$ edges connect the $m$ pairs of nodes of lower rank:
\[
P(r)=\begin{cases}
      0 & \text{if $r \leq m$}\\
      1 & \text{otherwise}\\
    \end{cases}       
\]
To interpolate between the two, we use a rational Bézier parametric curve, as illustrated in Fig. \ref{fig:bezierP}. The Bézier curve is defined by \textbf{two endpoints}, corresponding to the two points shared by both cumulative distributions: the points (0,0) and (L,m). The \textbf{control point} of the curve is (m,m). A weight $b$ allows controlling how close the curve is to each of the two extremes. If $b=0$, the curve corresponds to $\epsilon=0$, and $\epsilon(\lim_{b\to\infty})=1$. For convenience, we thus rescale the given parameter $\epsilon$ into $b$ as follows:
\[
b=\frac{\log(0.5)}{\log(1-\epsilon)}
\]
By convention, if $\epsilon=0$, $b=b^{max}$ and if $\epsilon=1$,$b=0$, with $b^{max}$ a large integer constant.

The function giving the probability of observing an edge between node pairs given their rank is defined by the derivative of the parametric Bézier curve (See Fig. \ref{fig:bezierP}).

The choice of the Bézier parametric curve arises as a natural solution to the problem as introduced in Fig. \ref{fig:sub1}: the curve for $\epsilon=0$ and $\epsilon=1$ are independent of the interpolation method, the chosen family function should thus propose a smooth interpolation between the two. The Bézier curve answers this problem in a convenient way, although other functions could be used.

\begin{figure}[h!]
\centering
\includegraphics[width=0.90\textwidth]{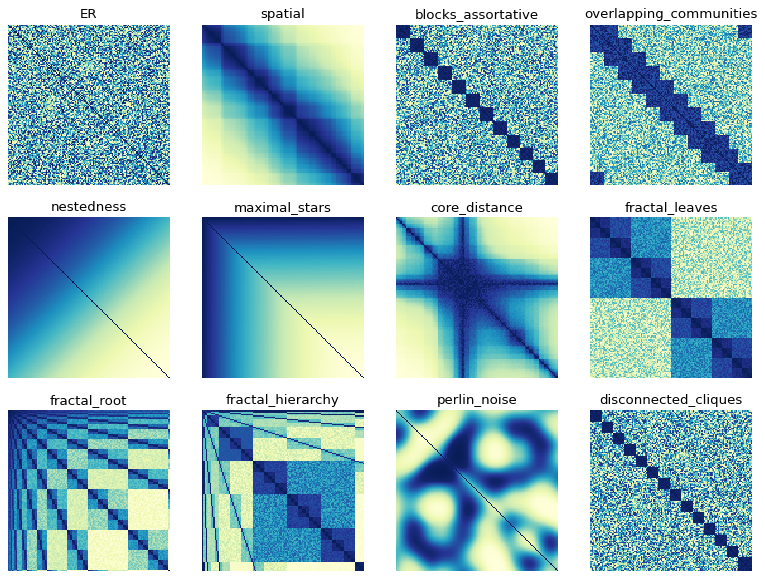}
\caption{The Structure Zoo. Matrix of node-pairs ranks for networks with 128 nodes. Darker colors correspond to lower ranks. For \textit{Disconnected cliques}, we set $m=128*8$. When involving spatial or clique positions, nodes are ordered according to this value.}

\label{fig:zoo}
\end{figure}

\section{\centering Structure Zoo}
\label{zoo}
To illustrate the expression power of the Structify-Net rank generation approach, we propose a collection of structures, available in the Python library under the name of Structure Zoo. This collection contains both classic structures widely found in the literature, together with original ones. Fig. \ref{fig:zoo} introduces matrix representations of the node-pair ranks of all structures in the zoo. The structures in the Zoo are only to be taken as examples, chosen arbitrarily among a few well-known structure types, and a few original ones. The main interest of Structify-Net is for the users to be able to specify their own structure with their own rank functions. The Zoo only represents a set of toy examples.

\subsection{Spatial structure}
Spatial structures are commonly found in the literature. Several versions of random graphs spatial models exist, for instance, the Waxman Graph (\cite{waxman1988routing}). More complex versions exist such as the Gravity model (\cite{cazabet2017enhancing}). A simple spatial structure can be easily implemented as a Rank model by using a cost Function,
\[
R'(u,v)=d(W_u,W_v)
\]
with $d(u,v)$ a notion of distance, such as Euclidean or Haversine distance. $W$ is a matrix such as $W_i$ is a vector representing the position of the node in a $d$ dimensional space. Positions can come from observed data, or be generated. In Fig. \ref{fig:zoo}, we attributed to nodes random positions in a 1-dimensional space. 

\subsection{Assortative block structure}
Community structure is one of the best-known types of organization of networks. A simple way to implement such a structure as a random graph generator is to use the stochastic Block Model (SBM), with a constraint of having an assortative structure, i.e., edges are more likely to be present between nodes affiliated to the same community than to nodes affiliated to different ones. A simple way to implement this as a rank model is using the following cost function:
\[
R'(u,v)=\begin{cases}
      0, & \text{if $B_u=B_v$}\\
      1, & \text{otherwise}
    \end{cases}       
\]
With $B$ the block affiliation vector, such as $B_i$ identifies the block to which node $i$ is affiliated with. Of course, many variants are possible, for instance, to take into account the size of blocks/communities.

\subsection{Overlapping Assortative Block structure}
A variant of the block structure allowing nodes to have multiple affiliations. There are numerous ways to model this situation. In the example provided here, we keep the same threshold cost function as for the non-overlapping case, extending it to multiple affiliations, i.e., we use the following cost function: 
\[
R'(u,v)=\begin{cases}
      0, & \text{if ($B_u \cap B_v) \neq \emptyset$}\\
      1, & \text{otherwise}
    \end{cases}       
\]
with $B_u$ the set of blocks to which node $u$ belongs. In Fig. \ref{fig:zoo}, each node belongs to exactly two communities and the affiliations are chosen such as each community has half of its nodes shared with another community c1 and the other half shared with another community c2.

\subsection{Block Structure: Disconnected Cliques}
Communities are often understood as sets of nodes that are strongly connected to each other and more weakly connected to the rest of the graph. A special case of extreme community structure can be set up by having only cliques, without links between them --- disconnected cliques. Keeping the same threshold cost function as for assortative blocks, we can find the value for community sizes such as obtaining the densest possible disconnected subgraphs for $\epsilon=0$, for a fixed $m$. Given the average degree $\widehat{k}=\frac{m}{2n}$, we want cliques to be of size $n_{c}\lceil \widehat{k} \rceil$. Because $n$ is not necessarily a multiple of $n_{c}$, we set the number of communities to $\lfloor \frac{n}{n_c} \rfloor$, and group the remaining nodes in an additional community. The already defined assortative block structure is then applied as usual with those blocks.

\subsection{Nested structure}
Nested network structures are well-known in some fields, such as ecology and economics  (\cite{mariani2019nestedness,alves2019nested}). A nested structure is a type of hierarchical structure, in which the properties/links of each entity are subsets of the properties/links of entities at a higher hierarchical level. We implement this as follows:
\[
R'(u,v)=u+v      
\]
Where $u,v$ are consecutive node indices in [1,n], and $u<v$

\subsection{Star structure}
Hubs are known to play an important role in many real networks. The hub-and-spoke structure is frequent both in human-designed infrastructure and in natural systems, forming patterns also known as \textit{stars}. One way to obtain such a structure is by using the following rank function: 
\[
R'(u,v)=u \times n + v      
\]
Where $u,v$ are consecutive node indices in [1,n], and $u<v$.

As seen in Fig. \ref{fig:zoo}, this rank function ranks first all the pairs of nodes including the node of ID 0, then all the pairs of nodes containing node ID 1 and another node with a larger ID, and so on and so forth. The structure therefore tends to create stars with nodes of low IDs in the center.

\subsection{Core Periphery}
Core periphery structure is another well-known type of organization for complex systems. This organization is often modeled using blocks, one block being the dense core, another block, internally sparse, representing the periphery, and the density between the two blocks is set at an intermediate value. To illustrate the flexibility of the Rank approach, we propose a \textit{soft-core} alternative, the \textit{coreness} dissolving progressively into a periphery. To do so, we consider nodes embedded into a latent space, as for the spatial structure ---random 1d positions in our example. The node-pair rank score is computed as the inverse of the product of 3 distances: the distances from both nodes to the center, and the distance between the two nodes. As a consequence, when two nodes belong to the center, they are very likely to be connected; two nodes far from the center are unlikely to be connected unless they are extremely close to each other. 
\[
R'(u,v)=d(W_u,W_v)d(W_u,\mathbf{0})d(W_v,\mathbf{0})
\]
with $\mathbf{0}$ the vector corresponding to the center of the space, i.e., the \textit{core} of the generated network.

Note that this is only an example of a rank function implementing a soft core, and one could imagine many variations of it.

\subsection{Perlin noise}
Perline noise (\cite{perlin2002improving}) is a type of gradient noise frequently used in computer graphics to create images with a realistic feel, such as textures and landscapes. We use it to generate an adjacency matrix, from the upper triangle of a 2d image of size (in pixels) $n \times n$. The $R'$ cost function is the black intensity of the pixel. In practice, Perlin noise tends to create continuous shapes of lower and higher values, with smooth transitions between the two (see Fig. \ref{fig:zoo}) for an example. Such a structure can be interpreted as a fuzzy version of a non-assortative SBM; with stronger relations between some groups of nodes and some other groups of nodes. Perlin noise has a parameter, called \textit{octaves}, allowing the addition of smaller-scale structures on top of each other.

\begin{figure}[h]
\centering
\begin{subfigure}{.5\textwidth}
  \centering
  \includegraphics[width=.9\linewidth]{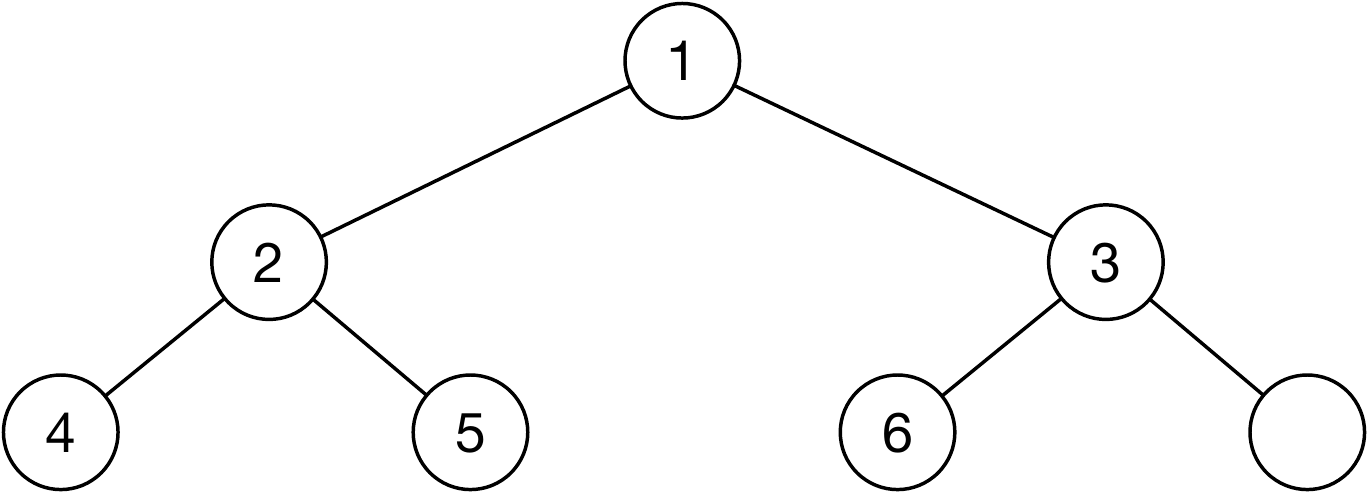}
  \caption{Fractal Root tree embedding}
\end{subfigure}%
\begin{subfigure}{.5\textwidth}
  \centering
  \includegraphics[width=.9\linewidth]{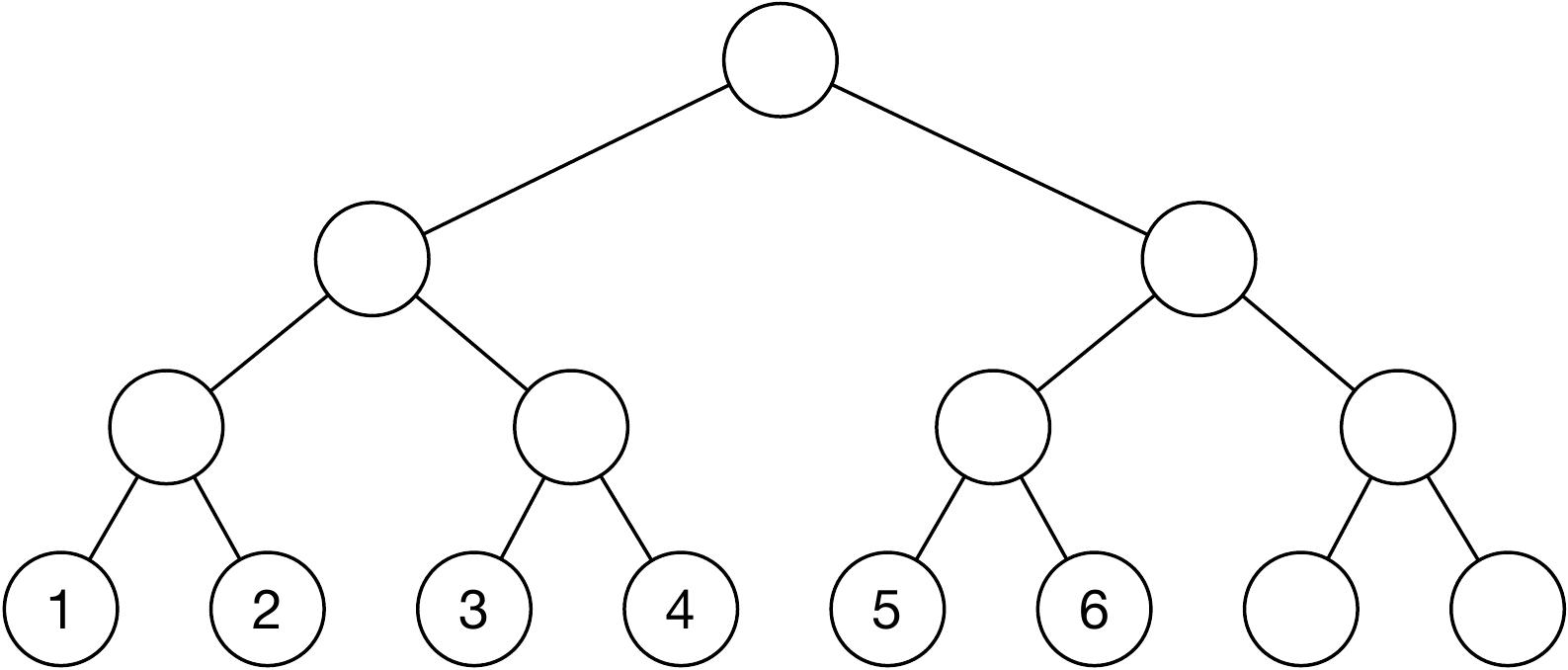}
  \caption{Fractal Leaves tree embedding}
\end{subfigure}
\caption{Two methods to create fractal structures by embedding nodes into complete binary trees. In the example, we embed 6 nodes. In the simpler case, the probability of observing a graph in the resulting graph is proportional to the distance between the nodes in the tree.}
\label{fig:tree_embedding}
\end{figure}

\subsection{Fractal structures}
To better illustrate the expression power of the Structify-Net structure definition, we propose three variations of what we call \textit{fractal structures}. The principle is to embed the nodes into a complete binary tree and to compute the rank scores based on distances on that binary tree. The purpose is to introduce heterogeneity among nodes, which can be used to create specific structures.

\subsubsection*{Fractal leaves}
In the fractal leaves structure, we create a complete binary tree such that the number of leaves is $n$ (Fig. \ref{fig:tree_embedding}). We embed nodes of the network on the leaves and use the distance between them in the graph as the cost function. This creates (see \ref{zoo}) a sort of \textit{Matryoshka doll}, hierarchical block structure, in which ---considering that edge probability decreases with distance without reaching zero--- small dense blocks are contained into larger, sparser blocks, recursively.

\subsubsection*{Fractal root}
In the fractal root structure, we embed nodes of the graph in a complete binary tree of size $n$, and define the cost function as the distance between nodes in the embedding tree (Fig. \ref{fig:tree_embedding}), $R'(u,v)=d^T(u,v)$, with $d^T$ the geodesic ---shortest path--- distance between the nodes in the embedding tree. This structure has similarities with the previous one, but introduces a particular role for some of the nodes: the root and nodes close to the role now occupy a central, pivotal role, since they are on the shortest paths between nodes on their rights and on their lefts. The network has both a hierarchy of blocks and a sort of central core composed of the nodes close to the root of the tree.

\subsubsection*{Fractal hierarchy}
The fractal principle and custom rank function can be used to create random networks with particular properties of interest. For instance, it has been pointed out in a seminal article(\cite{ravasz2003hierarchical}) that most real networks have a negative correlation between nodes' individual clustering coefficient and their degrees, while most network models do not reproduce this correlation. In the original article, a network model called the "hierarchical network model" has been introduced to generate networks with these properties. To reproduce the so-called \textit{hirarchical property} ---note that many other notions of hierarchical networks exist and that this is only the one introduced in (\cite{ravasz2003hierarchical}---  networks must have 1) heterogeneous degrees, 2) a high average clustering coefficient, and 3) a controlled relation between degrees and clustering coefficient. In the original article, such networks were created through an iterative deterministic algorithm, replacing graph parts with predefined subgraphs until reaching the target size. Instead, we propose here a rank-score approach, embedding nodes in a complete tree as in the fractal root embedding. However, we propose to use a ternary tree instead of a binary one, to increase local clustering. We then choose a score function such as: 1) leaves tend to have high clustering and low degree, and 2) root and other nodes in the higher levels tend to have high degrees and low clustering coefficients. The principle is thus to have a high probability of observing edges 1) between groups of nodes at the bottom of the tree if they have close common ancestors and 2) between nodes at the top of the tree and nodes at the bottom of the tree. The purpose of this example is to show that, by designing an appropriate rank function, one can obtain random graph generators such that the generated graphs have a property of interest.

The rank-score is thus defined as follows:
\[
R'(u,v)=\begin{cases}
      D(T_u,T_v), & \text{if $ANCESTOR(T_u,T_v$)}\\
      S(T_u,T_v), & \text{otherwise}
    \end{cases}  
\]
With $T_u$, the position of node $u$ in the embedding tree, $ANCESTOR$ a function such as
$ANCESTOR(u,v)=TRUE$ if $u$ is an ancestor of $v$ in the embedding tree. Functions $D$ and $S$ are scores capturing respectively a Descendent and Sibling similarity. We use:
\[
D(u,v)=min(h(T_u),h(T_v))+h(T)-(max(h(T_u),h(T_v))
\]
with $h(T)$ the global height of the tree and $h(u)$ the height of node $u$, such as $h(u)=0$ if $u$ is a leaf, and $h(u)=h(T)$ if $u$ is the root of the tree.
This function ranks first pairs of nodes that are far away in terms of tree levels, with a value of 0 between the root and the leaves.

\[
S(u,v)=\begin{cases}
      (d(T_u,T_v)-2)+h(T_u), & \text{if $h(T_u)==h(T_v)$}\\
      d(T_u,T_v)+h(T), & \text{otherwise}
    \end{cases} 
\]
with $d(u,v)$ the shortest path distance in the tree.

\subsection{Discussion on the structure zoo}
Structures introduced in this structure zoo are only a few examples of the infinite number of possibilities for structures that can be defined by cost functions. We stress that once such a cost function has been chosen, we are able to generate graphs following them, with a chosen number of nodes and edges. 

The structures we introduced allow the generation of synthetic networks without prior data, but one can perfectly define a cost function defined on node attributes, e.g., take a real network in which nodes are located in space, belong to known groups, and have other characteristic attributes, and define a structure by using a cost function taking all these attributes into account. Fig. \ref{fig:zoo} proposes a representation as matrices of all these structures on a network of 128 nodes and 1048 edges. 

\section{\centering Application: Swall World Structures}
One of the most famous properties of network structure is the so-called \textit{small world} property. Introduced in (\cite{watts1998collective}), it characterizes a network as being a small world if it has both 1) a high clustering coefficient --- significantly larger than in an ER random graph, 2) a short average distance --- of the same order of magnitude as in an ER random graph. This property, considered ubiquitous in real networks, has been reproduced in (\cite{watts1998collective}) by progressively adding randomness to a regular network, built such as the $n$ nodes are ordered in a circle, and each node is connected to its $\widehat{k}/2$ neighbors in both directions. The small world property emerges because, when we rewire edges at random, the average distance decreases faster than the clustering coefficient --- both being large in the regular network and low in the ER random graph.

We conduct an experiment to observe how other structures behave in terms of small-worldness when submitted to a similar experiment, i.e., starting with an archetypal structure, and adding noise progressively.

\begin{figure}[h]
\centering
\begin{subfigure}{.30\textwidth}
  \centering
  \includegraphics[width=.9\linewidth]{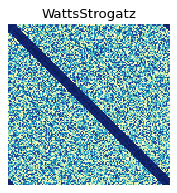}
  \caption{The proposed rank matrix. It has similarity with, e.g., the spatial one in Fig. \ref{fig:zoo}} 
  \label{fig:WS_matrix}
\end{subfigure}%
\hspace{1em}
\begin{subfigure}{.65\textwidth}
  \centering
  \includegraphics[width=.93\linewidth]{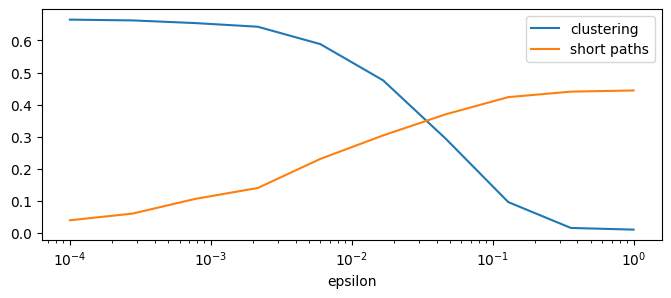}
  \caption{The small-world profile. As expected, the short path index increases significantly while the clustering coefficient remains close to the original value when increasing randomness}
  \label{fig:WSresults}
\end{subfigure}
\caption{Replicating the Watts-Strogatz experiment}
\label{fig:W}
\end{figure}



\subsection{Reproducing the Watts-Strogatz experiment}
\subsubsection*{Watts-Strogatz rank model}
To mimic the original small-world experiment, we define a rank-based structure using a cost function, parameterized by the number of nodes $n$ and the desired average node degree $k$.
\[
R'= \begin{cases}
      0, & \text{if $(v-u)\mod (n-k/2)<k/2$ }\\
      1, & \text{otherwise}
    \end{cases} 
\]
with $u,v$ node indices taken from $[0,..,n-1]$. The corresponding rank matrix is shown if Fig. \ref{fig:WS_matrix}

\subsubsection*{Scoring functions}
In the original article, clustering coefficients and average distances were expressed as a fraction of the value obtained for the regular structure. We cannot reuse this approach for multiple structures, having different starting values. Instead, for the clustering, we directly use the average clustering coefficient score, $CC(g)\in[0,1]$.

For the average distance, we propose a scaled value defined as:
\[
\widehat{\delta}(G)=\begin{cases}
      0, & \text{if $\frac{|\mathcal{G}(G)|}{n}\leq 0.9$}\\
      \frac{1}{1+\max(0,\widehat{d}(\mathcal{G}(G))-2)}, & \text{otherwise}
    \end{cases} 
\]
with $\mathcal{G}(G)$ the largest connected component of graph $G$, and $\widehat{d}(G)$ the average shortest path distance between nodes of graph $G$.

The property of this score is that $\widehat{\delta}\in[0,1]$, with $\widehat{\delta}=1$ if every node can reach any other node in two hops or less (e.g., a full star structure), and $\hat{\delta}$ decreases quickly as the average distance $\widehat{d}$ increases.

\subsubsection*{Watts-Strogatz experiment replication}
We use our setting to replicate an experiment similar in nature to the one in the original article. We used the same parameters, i.e., $n=1000,k=10$. We progressively add randomness, from a deterministic network to an ER random graph by varying parameter $\epsilon$ of the probability function. Fig \ref{fig:WSresults} shows results coherent with the original article: after adding some randomness, the short path index has increased significantly, while the clustering coefficient still remains close to its value for the deterministic network.


\subsection{Small-World profiles for other structures}

\begin{figure}[h]
\centering
\includegraphics[width=0.8\textwidth]{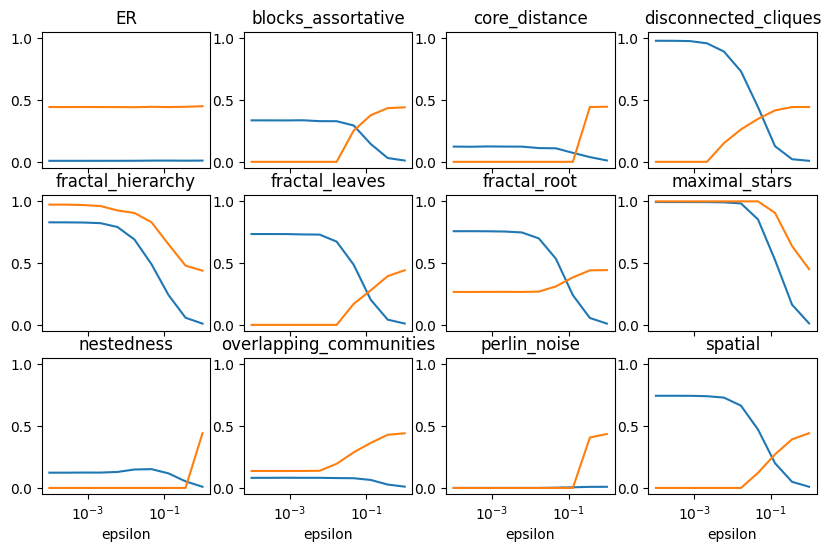}
\caption{Small-World profiles of networks generated by generators of the \textit{structure zoo}. \textbf{Blue} is clustering, \textbf{orange} is the short path index (lower values correspond to longer paths). We observe \textit{super-small-world}, e.g., fractal-hierarchy, \textit{anti-small-world}, e.g., nestedness, profiles similar to the Watts-Strogatz network, e.g., spatial or blocks assortative.}
\label{fig:WS_all}
\end{figure}

We can apply the same process to the other structures defined in our structure zoo, with the same number of nodes and edges. In Fig. \ref{fig:WS_all}, we observe a wide variety of behaviors. 
\begin{itemize}
    \item Fractal-hierarchy and maximal-star structures display a \textit{super-small-world} behavior, having both short paths and high clustering coefficients. This can be easily understood: their hierarchical nature creates a giant hub maximizing reachability. Fractal-hierarchy is designed such as most nodes of low degree have a high clustering coefficient, due to a local structure. On the contrary, in maximal stars, most nodes are connected only to a few hubs, but since those hubs are connected to each other, they also have a high clustering coefficient.
    \item Nested, overlapping communities, and Perlin noise seems, on the contrary, to be \textit{anti-small-world}, with both a low clustering coefficient and long paths. Again, this is due to different factors. For instance, the nestedness and Perlin noise concentrate so many edge probability between a small subset of nodes, that many nodes are disconnected, leading to the absence of a giant component ---thus to an infinite average distance. The overlapping community, instead, is created in a way that makes it look like the original Watts-Strogatz circular structure, as can be observed in Fig. \ref{fig:zoo}. Its low clustering coefficient comes from 1)many nodes not having a degree of at least 2, or 2)structures being too large compared to the number of edges, so that the probability of forming triangles is low.
    \item Other structures tend to follow a pattern roughly similar to the original article, with more or less pronounced profiles. In some cases, the short distance score remains at zero until reaching a certain amount of noise, due to the absence of a giant component.
\end{itemize}

\section{Working With Node Attributes}
In the previous section, we have seen an experiment in random networks without metadata/attributes, i.e., with interchangeable nodes. However, as mentioned in the model definition, it is possible to use any node information in the rank function. In this section, we provide two simple motivational examples.

Let's imagine working with a dataset such as a global air transport network. Nodes correspond to airports and are identified by their name, location, and country. Edges might correspond to having at least one direct flight between two airports. To study a property of such a network, for instance, its clustering, average shortest path, or properties of a virus diffusion on that network, one would usually compare that property with its value in networks generated by a reference null model, such as ER or Configuration model. Another possibility offered by Structify-net is to use the node attributes. For instance, we can consider only the airport positions, and rank node pairs according to the distance between them in the dataset, which corresponds to using the \textit{Spatial structure} defined in the Zoo, using metadata instead of random positions. Alternatively, we could use the country information as metadata to the \textit{Assortative Block Structure} of the Zoo. Of course, it could be possible to use already-existing methods such as block models or spatial models to do the same. Structify-Net simply offers a convenient way to do it, for any model, just by providing the appropriate function.

But it is of course possible to go far beyond simple blocks or latent space, and to propose a custom ranking function based on the case study. For instance, one can use machine learning to design a ranking function retaining complex properties from an observed graph. In the airport example, we could use a classification algorithm such as logistic regression or a decision tree to learn how likely it is to observe an edge between two nodes given their properties. From the observed network, we would extract a set of training examples $\{distance,sameCountry \in \{0,1\},edge \in \{0,1\}\}$, and train a classifier, which can then assign a class probability to each node-pair. This probability would not however be usable directly to create null models preserving the number of edges. Instead, we can use this probability as a rank function, and generate random graphs with a chosen number of edges using Structify-Net. The structure produced will preserve some of the properties related to attributes of the original graph ---probably, a higher tendency of airports to be connected if they are located in the same country and if they are close in space.

\section{\centering Python library}
An important aspect of such a generator is to allow other researchers to use it for their own needs, whether it be to generate networks according to a structure described in the structure zoo, or to define their own. We thus release with this paper a pip installable python library \cite{structify}, together with its documentation\footnote{\url{https://structify-net.readthedocs.io/en/latest/}}.
For convenience, the library is compatible with Networkx (\cite{hagberg2008exploring}). Obtaining a rank model corresponding to one of those defined in the zoo, such as the nested structure, is as simple as calling it:

\begin{lstlisting}[language=python]
import structify_net.zoo as zoo
n=128
rank_model = zoo.sort_nestedness(n)

\end{lstlisting}

Generating a network as a Networkx object from it is straightforward:

\begin{lstlisting}[language=Python]
import structify_net.zoo as zoo
n=128
m=512
generator = zoo.sort_nestedness(n).get_generator(epsilon=0.5,m=m)
g_generated = generator.generate()
\end{lstlisting}

One can also define a custom structure by providing a rank-score function:

\begin{lstlisting}[language=Python]
import structify_net as stn
n=128
m=512

def R_nestedness(u,v,_):
    return u+v
rank_nested = stn.Rank_model(n,R_nestedness)
g = rank_nested.generate_graph(epsilon=0.1,m=m)
\end{lstlisting}

The library allows easy plotting of the rank-score matrices and node-pair probability matrices, and more generally reproduces all the content of the current article.

\section{Related Works}
\label{relatedWork}

Many works can be found in the literature on the generation of random graphs. A complete survey is beyond the scope of this paper; we will nevertheless briefly introduce in this section the most common random graph models and existing software for random graph generation.

We can make a distinction between generative models that are designed for model parameter inference, and those that are not. Models designed for inference are usually controlled by a limited number of parameters, and have some appropriate statistical properties allowing to infer parameter values to match a specific observed graph or series of graphs(e.g., SBM or latent-space models). On the contrary, generative-only models are designed to generate graphs with some specific properties in order to use them in downstream tasks (e.g., LFR or Waxman graphs). Structify-Net, although defined as an edge-independent random graph model, rather belongs to the second category, since it is not designed for parameter inference.

\subsection{Common Random Network Models}
The simplest way to generate random graphs is certainly the Erdos-Renyi (ER) random graph model, that we already introduced. ER random graphs are fully homogeneous and do not have any particular structure; but a controlled expected size. Configuration models, in particular the Chung-Lu version, allow the generation of graphs of controlled size without mesoscopic organization but preserving the nodes' degrees.

\subsubsection*{Block structures}
Stochastic Block Models (SBM) define random graph models with block structures. In their simpler form, they are defined by two sets of attributes: a vector defining the block to which each node belongs, and a matrix defining the number of edges between each pair of blocks. They exist in various flavors, the canonical version being edge-independent (\cite{snijders1997estimation}), while microcanonical versions (\cite{peixoto2017nonparametric}) are not. These models also exist with or without node degree preservation.
The literature on SBM mostly focuses on inference, but SBM can naturally be used to generate random networks, either by choosing model parameters or by using those obtained after inference. A popular way to set manually the parameters for custom structure generation is to fix a number of blocks and a number of nodes in each block, then to choose an internal edge probability $p_{in}$ and an external probability $p_{out}$. Usually, $p_{in}>p_{out}$, thus defining assortative blocks. Arbitrary block structures, with different block sizes or non-assortative structures can be defined by setting the parameters accordingly.

Multiple variants of block models exist, such as overlapping SBM(\cite{latouche2009overlapping}) or hierarchical ones (\cite{schaub2023hierarchical}). Block models can also be used to generate core-periphery structures, typically by setting one core block and one or several peripheral blocks. This structure however cannot generate other types of possible core-periphery structures, such as a continuous change between core and periphery.

A popular random graph generator with community structure is the LFR Benchmark(\cite{lancichinetti2008benchmark}). Not designed for inference, it allows the generation of networks with realistic properties with a limited number of parameters, thus in a more convenient way than with manually initialized SBM. A more recent variant solving some of the problems of LFR is the ABCD random graph generator(\cite{kaminski2021artificial}).

\subsubsection*{Latent space structure}
Various models exist to generate random graphs in which nodes are embedded into a space, the probability of observing an edge depending on the distance between them. Among popular examples, we can cite Random Geometric Graphs (RGG, \cite{dall2002random}), in which nodes are connected if their distance is below a parameter, and Waxman Graphs (\cite{waxman1988routing}), in which edge probability decreases exponentially with the distance. The gravity model(\cite{wojahn2001airline}) is an alternative in which the probability of observing an edge depends both on nodes' degrees and on a deterrence function controlling the influence of distance on edge probability. The parameters, in particular the deterrence functino, can also be inferred to fit a given observed network (\cite{cazabet2017enhancing}). Latent spaces are not limited to geographical space, and models have been proposed for the inference of social spaces, for instance (\cite{hoff2002latent}).

A model related both to SBM and to spatial models is the Random Dot Product Graph (RDPG) model (\cite{young2007random}). Nodes are characterized by a vector defining their positions in a latent space, and the probability of observing an edge between nodes is given as the dot product between their vectors.

Some authors consider instead that networks are better represented in hyperbolic space, leading to the proposition of Hyperbolic random graph generators (\cite{aldecoa2015hyperbolic}).



\subsubsection*{Homophily}
Other generators model edge probabilities depending on the nodes' attributes. They allow to analyze the interplay between similarities in structure (e.g., common friends in social networks) and similarities in node attributes (\cite{asikainen2020cumulative}), or to investigate mechanisms of non-structural closures such as the formation of links between nodes having similar attributes that do not share common neighbors, as a base of node homophily (\cite{murase2019structural}).

\subsubsection*{Generic random graph models}

A difference between models introduced until then in this section and Structify-Net is the restriction to a single type of network structure. Since Structify-Net accepts any node-pair ranking function as input, it allows the generation of block, spatial, but also other types of structures such as core-periphery, nestedness, etc.

Another family of highly expressive random graph models is the Exponential Random Graph Model family (ERGM, \cite{lusher2013exponential}). ERGMs define the probability of observing a given graph $G$ as $P(G) = \frac{exp(\Theta \cdot X(G))}{c(\Theta)}$, with $\Theta$ a vector of network parameters, $X(G)$ network characteristics, including for instance the number of triangles or node properties, and $c(\Theta)$ a normalizing constant ensuring that the sum of $P(G)$ for all $G$ is equal to 1. ERGMs are mostly used in the context of model inference, and allow in theory to model non-independent edges, e.g., taking into account a triangle closure propension. However, due to the computational complexity, this approach is limited to small graphs.

Finally, an approach sharing some similarities with our framework is the \textit{graphon} (\cite{glasscock2015graphon}), contraction of \textit{graph function}, first introduced in (\cite{lovasz2006rank}). A \textit{graphon} can be defined (\cite{orbanz2014bayesian}) as a bivariate function $W : [0, 1]^2 \rightarrow [0, 1]$. That function returns an edge probability for each pair of nodes, based on a node latent variable. Graphons were first introduced mostly as theoretical objects, in the context of sequences of large, dense graphs. More recently, works have focused on the inference of this nonparametric model, as smooth-graphons (\cite{sischka2022based}) or combined with an SBM approach (\cite{sischka2022stochastic,orbanz2014bayesian}). While graphons share the principle of using a function to characterize the network structure with our approaches, they are part of a very different literature. Graphons are more generic, so much so that SBM, spatial, and nearly all latent-variable-based statistical models can be considered a special case of graphons. The literature on the topic focuses on inference problems, and notions such as node-pair ranking or randomness parameters are not present. 

\vspace{0.5cm}

The Structify-net framework is aimed to play a different role compared with all methods introduced in this section. ERGM and Graphons are families of models, designed for model inference rather than network generation. They are so general that they do not offer much help to define a particular structure, and are used in general in a restricted context, for instance with block-approximations for graphons, or imposed number of triangles for ERGMs. 
SBM, gravity models and configuration models are on the contrary more specific than structify-net, focusing on a single type of network structure. Furthermore, they are often used in the context of model inference. 
On the contrary, other models such as LFR benchmark or Waxman graphs are designed, as Structify-Net, to generate networks with controlled properties, but they also focus on one specific type of structure. Our contribution thus occupies an original position in the scientific landscape on random graphs: it is designed for the generation of random graphs and not the inference task, it is more flexible than LFR or SBM, and offers a more convenient way to generate graphs of controlled properties compared with ERGM or Graphons.

\subsection{Software}
Several easy-to-use libraries propose to generate networks with blocks, following the Stochastic Block Model approach. Among the most popular, we can cite networkx  (\cite{hagberg2008exploring}) and iGraph  (\cite{csardi2006igraph}), which include an SBM generation function allowing to define blocks of arbitrary sizes, arbitrary probabilities of observing edges between them, and then generate a graph accordingly. More advanced functions are proposed in the graph-tool (\cite{peixoto_graph-tool_2014}) library, allowing to generate \textit{microcanonical}, degree-preserving versions, and several other variants of block structures. 

The same libraries offer, under the name of \textit{geometric models}, some possibilities for spatially constrained network structures. Most of these methods, however, do not allow setting the number of edges, since they instead require setting a threshold below which edges exist, or using an a priori function (Waxman random graph). Only the \textit{k-nearest neighbors} method allows one to choose the number of edges among multiples of the number of nodes, but it is a deterministic generator. These libraries also contain other types of network generators with non-random structures, such as lattices, or networks defined by a process such as the forest-fire model.

Another notable network generator is the LFR benchmark, which is implemented in networkx in a simplified version, or available as a standalone code to have access to all its settings. Software to work with graphons is much more scarce; we found an R library implementing graphon inference and graphon random graph generation \footnote{\url{https://cran.r-project.org/web/packages/graphon/index.html}}; a recent python code exists also, although not in the form of a documented library \footnote{\url{https://github.com/BenjaminSischka/GraphonPy}}. These libraries however have nothing to see, in term of usage of capabilities, with Structify-Net. They are designed for completely different purposes. The reference library for working with ERGM is the R package \lstinline{ergm}(\cite{hunter2008ergm})\footnote{\url{https://gvegayon.github.io/appliedsnar/the-ergm-package.html}}, focusing on model inference.

\section{\centering Discussion}
This article introduced a new method to generate random networks with a customizable network structure, and a target number of nodes and edges, while controlling the amount of randomness. To the best of our knowledge, this is the first random network generator allowing to do so.
We think that having such a generator opens doors to new research directions in network science, for studying the properties of networks with some particular structures ---as we have done in the experimental section, or as a reference model for observed graphs, to name a few.

Moreover, one of the main strengths of the generator is its ability to control situations where a process/rule of the structural organization (expressed by a pair-node rank) can be mixed with "unknown" random processes (expressed by $\epsilon$); thus, among the observed edges, some of them strictly follow the structural constraints imposed by the rank, and some of them can go beyond the explanation of such constraints.
The possibility to analyze this mix ---between edges driven by the organization and randomness--- is quite important, especially given that network behaviors like small-worldness or homophily can be better explained when randomness is added to a rule of connection (\cite{talaga2020}).

Regarding the analysis of such network behaviors, another strength of the generator is the possibility to exploit the properties of nodes when defining a rank, thus including elements representing, in principle, physical position, political opinions on a spectrum, gender, or even degree.
We focused here on the distance between vectors of nodes' positions in a $d$ dimensional space for building a spatial structure (cf. "Structure Zoo", 2.1).
We also focused on the affiliation to the same group for building an assortative block structure (cf. "Structure Zoo", 2.2).
Similarities between such structures lead us to acknowledge the significance of incorporating node metadata/attributes to generalize a wide variety of network behaviors.
We focused here on analyzing the small-world property, but the same can be applied to other behaviors.
In principle, behaviors like homophily (\cite{mcpherson2001birds}), could be described just as a particular case of either a spatial organization (if attributes are numerical) or of an assortative block structure (if attributes are categorical).

\subsection{Limits and future work}
The main limit of the current work is scalability: node-pair ranks and probability matrices are dense matrices, which can be memory-demanding for large graphs. The generation process also requires an independent random draw for each node-pair. These limits could be overcome in future work. Another limit is that network structures in which probabilities of observing an edge between a pair of nodes are not independent of adjacent ones ---for instance, to generate random networks of a specific size and a specific clustering coefficient--- cannot be expressed by a rank-structure. Another limit compared with some other random graph models such as SBM or RDPG is the absence of an efficient inferential solution. In particular, without any constraint on the node-pair order, an inferential approach would always find a trivial uninformative solution in which all connected node-pairs are ranked first. Note however that, since it is possible to compute the probability to obtain a given graph for a set of parameters, it could be possible in theory to use maximal-likelihood inference on a subset of the models, e.g., by fixing the amount of randomness $\epsilon$, and constraining the domain of acceptable rank functions.

In future work, we plan to compare the properties of real-world networks with those of the synthetic ones generated from structures such as those of the zoo. Having such a variety of possible structures, we expect to be able to characterize real networks, by observing similarities and differences with the synthetic ones, e.g., a real network might have a clustering coefficient and an average distance compatible with the Watts-Strogatz network, but differ in degree heterogeneity and robustness, while another synthetic network might have more similar properties in all those aspects. In particular, we will investigate the role of randomness, to test the original idea of the Watts and Strogatz small-world definition, i.e., that randomness is at the source of complex networks' properties.


\section*{\centering Fundings}
This work is supported by the European Union – Horizon 2020 Program under the scheme “INFRAIA-01- 2018-2019 – Integrating Activities for Advanced Communities”, Grant Agreement n.871042, “SoBigData++: European Integrated Infrastructure for Social Mining and Big Data Analytics” (http://www.sobigdata.eu).

This project was partly founded by BITUNAM grant ANR-18-CE23-0004.

\section*{\centering Conflict of interest disclosure}
The authors declare that they comply with the PCI rule of having no financial conflicts of interest in relation to the content of the article.

\section*{\centering Data, script, code, and supplementary information availability}
Scripts and codes are available online as a documented, pip-installable Python library.
\begin{itemize}
    \item GitHub (\url{https://github.com/Yquetzal/structify_net}
    \item DOI: \url{https://doi.org/10.5281/zenodo.7966895}
    \item Documentation: \url{https://structify-net.readthedocs.io/en/latest/}
\end{itemize}.

\titleformat*{\section}{\bfseries\Large\centering}

\printbibliography[notcategory=ignore]

\end{document}